# Collecting MIB Data from Network Managed by SNMP using Multi Mobile Agents

Nisreen Madi
*Computer Science Department*
*Princess Sumaya University for Technology*
Amman, Jordan
nisreen.madi@gmail.com

Mouhammd Al-kasassbeh
*Computer Science Department*
*Princess Sumaya University for Technology*
Amman, Jordan
malkasasbeh@gmail.com

*Abstract*— Network anomalies are destructive to networks. Intrusion detection systems monitor network component behavior to detect unusual activity (i.e., possible threats). Application-layer Simple Network Management Protocol (SNMP) has been used for decades via TCP/IP protocol to manage network devices. Raw data security evaluation in intrusion detection incurs latency in detection. Management Information Base (MIB) combined with SNMP is a solution for this, the traditional approach of SNMP is centralized. Thus, rendering it unreliable and non-adaptive to network changes when it comes to distributed network. In distributed network, using single or multiple light Mobile Agents are an optimal solution for data gathering as they can move from one source node to another, executing naturally at each. This helps complete tasks without increasing the network overheads, and contributes to decreasing latency.

This paper focuses on finding the optimal number of mobile agents to complete the data retrieval task with the minimum routing time, without affecting the network bandwidth, to solve the Simple Network Management Protocol-Management Information Base centralization issue, and enhance detection time. Two types of agents are used in this paper; link agent for discovering the network, and data agents for MIB gathering. The link agent runs in the home node to discover the network and define nodes' connectivity. Then, network is partitioned based on its execution time. Single mobile agent is sent to each partition to complete MIB retrieval task. This approach aims to finish MIB retrieval task with minimum routing time and keeps generating of mobile agents under control to maintain optimal network bandwidth. Our approach are enhancement on two approaches were proposed in previous studies in the same filed this paper will present details on each approach and conduct a comparison regarding number of agents used to gather MIB data and the time needed to complete the gathering task

## I. INTRODUCTION

Distributed Network (DN) has been studied widely in numerous researches due to its widespread applications in the military and civilian domains, whether in environmental monitoring, such as pressure and temperature, or in tracking objects, such as object locators (Holger & Willig, 2007). The client/server model has been implemented widely in DN. The client can request a service execution from the server, which deploys a group of methods to access resources to perform the requested service in response to the client request (Fuggetta, Picco, & Vigna, 1998).

Any network can be exposed to different attacks which could affect the data, the process, or the system itself. Intrusion Detection System (IDS) monitors the network and detects any network anomaly or attack, by comparing the occurring activity with the node normal behavior, which helps detect any malicious activities or attacks that could affect data integrity, confidentiality, or availability (Al-Kasassbeh, Al-Naymat, & Al-Hawari, 2016). Simple Network Management Protocol (SNMP) is commonly used for this purpose. It is an application layer protocol used to manage the network devices. It runs over the user Datagram Protocol and collects information and configuration from network devices (Thottan & Ji, 2003).

The information gathered from the network devices such as routers, hubs, and printers etc. could be used in analyzing the devices behavior and detecting any attacks when they occur (Astuto, Mendonça, Nguyen, Obraczka, & Turletti, 2014).

Management Information Base (MIB) was combined with SNMP to manage networks and detecting anomalies by minimizing the delay in analyzing devices' behaviors and detecting anomalies when they occur. MIB is defined as a tree structured database used for managing a set of network objects. Each device has an MIB variable that can be effected somehow when any anomaly occurs (WhatIs.com, 2018). For network management, many data objects are maintained by MIB, such as link status, equipment status, system data, and communication status etc. (Du, Li, & Chang, 2003).

The client/server model is implemented in most IP management networks. Since SNMP is considered as the core protocol for most networks managed by MIB, it shares the advantages and disadvantages of the client/server model. We can say that SNMP consists of a manager application, static SNMP agent (defined as an interface between the legacy system and the manager application), and physical resources (Damianos, Ysekouras, & Anagnostopoulos, 2009). Within SNMP, the manager application is performed as the client, while the SNMP agent is performed as the Distributed server. Finally, the managed objects are considered as the physical resources (Liotta, Pavlou, & Knight, 2002).

There are spectacular disadvantages for using client/server model though it has been spread widely. Its processing centers need higher energy and computation time, which dramatically reduces the lifespan of nodes, especially in homogeneous and autonomic networks. Furthermore, DN contains many clients, so when the data gathered by nodes moves from the clients to the server, it may require larger network bandwidth than the available resources can support, causing poor performance (Xu & Qi, 2008).

As a solution, Mobile Agent (MA) has been used as an optimal solution to retrieve data in DN due to its nature as software able to migrate between nodes and continue execution naturally (Konstantantopoulos, Mpitziopoulos, Gavalas, & Pantziou, 2011). Each MA has its own capabilities that determine the type of tasks it can handle. Agents are able to delegate a task to another agent, or clone themselves if they cannot perform a task autonomous (Aridor & Oshima, 1998). In a controlled environment, MAs can also minimize the traffic, overcome network latencies, and boost the robustness of distributed applications (Shehory, Sycara, Chalasani, & Jha, 1998).

Mobile Agent Planning (MAP) is necessary to minimize routing costs and plan agents' paths to complete tasks in the optimum way, to retrieve data from MIB for observing the distributed network with the minimum cost, time, and energy (Baek, Yeo, Kim, & Yeom, 2001). This is the main area explored in the current paper.

Since SNMP-MIB traditional methods are centralized, we used an MA solution to collect MIB data. The time needed to collect MIB data is very important in enhancing the intrusion detection efficiency. Thus, we aim to find the optimal number of MAs to complete MIB data retrieval task with minimum routing time, without increasing network bandwidth.

MAP's most advantageous performance factors include the number of agents used and execution time. Fewer agents mean less network traffic, and thus less bandwidth consumption, and regardless of the number of agents, the execution time should be minimal.

Our motivation in this work is defining the optimal number of agents to retrieve data from MIB to monitor the distributed network from the nodes with minimum routing time, and without increasing the network traffic.

The remaining of this paper is organized as follows. Section II reviews literature related to this study, while section III presents the research methodology and proposed approach. Section IV presents the experimental results and main findings, while section V concludes this paper, finally, section VI demonstrate the reference.

## II. RELATED WORK

### A. SNMP and MIB

Damianos, Ysekouras, and Anagnostopoulos (2009) designed a complete MAP research prototype that addressed security and fault tolerance issues. It was implemented with java and optimized for systems and network management applications. Their prototype's main objective was adopting modular MA architecture, which eases the reusability of the code, along with adding new services or modifying the existing ones. The prototype consists of the following components:

- Manager responsible for MA code.
- MAs that can migrate between the managed objects and collect information.
- Mobile agent server (MAS) responsible for receiving MAs and extending the interface to physical objects.
- Mobile Agent Generator (MAG) responsible for automatically creating and deploying agents.

They used NS2 to evaluate their prototype on a large network scale. The results showed that the prototype is approximately symmetrical to the size of the managed network.

Pagurek, Wang, and White (2000) demonstrated the need for implementing MA in the management network, extending an existing mobile agent framework used in the management network domain. The implemented design used the Distributed Protocol Interface (DPI), and a proposed RDPI protocol extended from DPI for enhancing the interaction between SNMP agents and MAs.

Al-Kasassbeh and Adda (2008) presented the centralized paradigm of SNMP and its disadvantages. The traditional SNMP paradigm stores management information in the management information base (MIB), and each node in the system will has its own MIB, with information being obtained by using management protocols. This approach is suitable for applications that have a restricted need for distribution control. This approach has many disadvantages, such as inadequate scalability, flexibility, and availability for distributed network with a large number of nodes. This paper presents how to assess MAs and client server paradigm performance by building an analytical framework. Their proposed framework is based on adaptive intelligent MA, which is a combination of classic MA and CS. Two approaches are proposed to poll the data from the domain, Accumulative Model and Interactive Model, which are explored later.

Wittner, Helvik, and Hoepler (2000) addressed the importance of fault management, including the isolation, discovery, and fixing of problems, and how efficiency is critical in recovering from network faults that may appear in the cycle. They also addressed the scalability limitation of centralized network management, especially when considering transferring bulk network monitoring data.

Al-Kasassbeh, Al-Naymat, and Al-Hawari (2016) described types of DoS attacks and their network impacts.

They also demonstrated challenges facing the IDS, such as the lack of real-life datasets to be used in anomaly detection. Most data used for this purpose comprise results from simulation approaches, which undermines the accuracy of the outcomes as simulated conditions do not genuinely reflect the full extent of real-world scenarios of intrusion or anomaly. The paper defined the most important requirement to generate a dataset as overcoming the aforementioned shortcomings. It demonstrated a real-life testing dataset for attacks' traffic, using SNMP-MIB statistical data gathered from the designed dataset. The paper also provided 4998 records containing 34 MIB variables that can be used to test the presented IDS.

All of those previous researches mentioned how MAs could be integrated to decentralize network management systems and how to add or edit services that enhance the security management, but none talked about how to enhance the speed of gathering the data to enhance detection and make it more efficient, as presented in this research.

B.   *Single Mobile Agent Planning*

Baek, Yeo, Kim, and Yeom (2001) focused on two performance factors in MAP: the number of MAs and the total routing time consumed by agents to complete a task in the distribution system. They proposed two heuristic algorithms, BYKY1 and BYKY2, whereby nodes should be organized in descending order, and the network was divided such that each part's execution time did not exceed the execution time of the first node in the ordered list. Finally, the shortest path for each part was determined using TSP, and then the MAs were sent from the home node to each part.

Building on the previous study's MAP algorithm, Baek, Kim, and Yeom (2001) took a third performance factor into consideration: time constraints. The travelling time of MA is important due to the ability to cut off the number of agents used to complete a task if we minimize the agent's travelling time. They used a 2OPT–TSP to optimize agents' local network path. This method considerably minimizes the number of agents along with the routing cost.

These papers partitioned the entire network without checking nodes' connectivity, so the agents could pass through a disconnected agent unnecessarily (Baek et al., 2001; Baek, Kim, & Yeom, 2001). Additionally, they did not allow any cloning, which would have enabled the MA to delegate the task to another agent when overloaded.

Moizumi (1999) showcased a number of Travelling Agents' Problems that occur in MA information retrieval and data mining applications. The planning problems related to the best sequence of sites to be visited, which minimizes the expected time needed to complete a task. The thesis talked about both single and multiple agent problems. A polynomial and semi-polynomial algorithm was successfully developed for such problems along with implementing a planning system that used these algorithms.

This model requires some network statistics to be known, such as link bandwidth, site density, and latencies. Referring to these statistics, agents can find the best path with the minimum time and cost to a specific location. Moizumi (1999) stated that the TAP is NP-complete in its general formula. Clustering the sites makes latencies between them approximately constant, which makes the TAP less complex and decreases its computation to polynomial time.

Chen, Kwon, Yuan, Choi, and Leung (2006) proposed an algorithm called Mobile Agent Directed Diffusion (MADD) using Local Closest First (LCF) with one change: they start with the furthest node. In this algorithm, nodes take an active part in the itinerary planning of the MA, and each node has to maintain a secure entry table that should be inserted in the next node visited, based on the task. Since the nodes have a limited memory, a memory issue is raised for each node that stores the table. After finding the nodes that should be visited to complete a task, a single agent should visit all the nodes, so if there are a huge number of nodes for the same task, the agent will not be effective enough to visit all nodes and collect data.

C.   *Multi Mobile Agent Planning*

Qi and Wang (2001) developed a method to find the optimal path for MAs to achieve the integration task while consuming the minimum power and time. The paper stated that dynamic path planning is more flexible and can adapt to changes in the environment, but it consumes more computational time and power than static planning. The optimal method should be applied before dispatching the agents and giving them the liberty to return to the dispatcher without completing the trip, once the results' accuracy reaches the threshold required for task completion.

The main focus of Qi and Wang (2001) was finding the optimal path with minimum power and time, without taking the routing cost into consideration. The main idea is to find the optimal path and make the agents communicate with each other, so if one completes a task the agents' trip should be terminated. However, the algorithm did not take the density or distance between nodes into consideration, and it followed the classic critical path in planning. Moizumi (1999) stated that in Travelling Multiple Agents Planning, it is assumed that agents communicate with each other, and once one agent completes the task the others will stop execution. The problem of TMAP is to complete task with the minimum time (i.e. the system aims to minimize the expected time).

Prapulla, Chandra, Mudakavi, Shobha, and Thanuja (2016) stated in their paper that multi agent planning is a key technology used for optimized the energy consumption in WSN. The MAs used were link agent and data agent. Link agent is responsible for new nodes added to the network, and connectivity and disconnection issues in the network. Data agent is part of the information transferring process between nodes. They used Farthest Node First Nearest Node Next for implementation. They started by clustering the network, then they used the link agent to test network connectivity. Moreover, the heads of the clusters determine what nodes will be assigned to each agent, taking into consideration the amount of its data and the distance between it and the cluster head. In addition to taking care of the distance, the algorithm also enables the MA planning to be in the same order as the nodes assigned to each MA. Clustering in this context is not dependent on network changes, which makes the algorithm less adaptive to changes such as the number of nodes available and the density. While Prapulla et al. (2016) started by clustering the network then using the link agent to test nodes' connectivity, in our approach we read the nodes' connectivity first, then partition the network based on a shortest latency graph and the density of the available nodes.

Qadori, Zulkarnain, Hanapi, and Subramaniam (2017) reviewed how multi MA planning overcomes single MA planning faults in terms of task delays and agent size. Also, it mentioned a number of existing algorithms used in MMAP that addressed a number of critical issues, such as defining the optimal number of MAs, source nodes grouping, and finding the optimal path for agents for data retrieving task. The review stated that recent algorithms are taking MMAP parameters into consideration while ignoring other parameters, and none of the reviewed algorithms took the security of data gathered by MAs into account.

### III. METHODOLOGY AND PROPOSED APPROACH

This chapter presents the description of the proposed methodology of the Multi Mobile Agent Planning approach.

The main concerns related to MAs are protecting agents from malicious hosts, or protecting hosts from malicious agents. Our algorithm aims to implement authentication and authorization policies to ensure trust between the host and the agent, and to overcome the aforementioned issues.

We assumed that each node in the network has 1KB of MIB data to be retrieved; the MIB data size is very small, thus we excluded the delay at each node in our calculation, since it is the same at each node. We used two types of agents: link agents and data agents. Link agents are run as a background service to retrieve connected nodes, while data agents are used for retrieving data from nodes.

Figure 3.1 shows a sample of a network containing 10 nodes managed by SNMP, displaying the weight on the paths between nodes. This sample is used to clarify the algorithm steps.

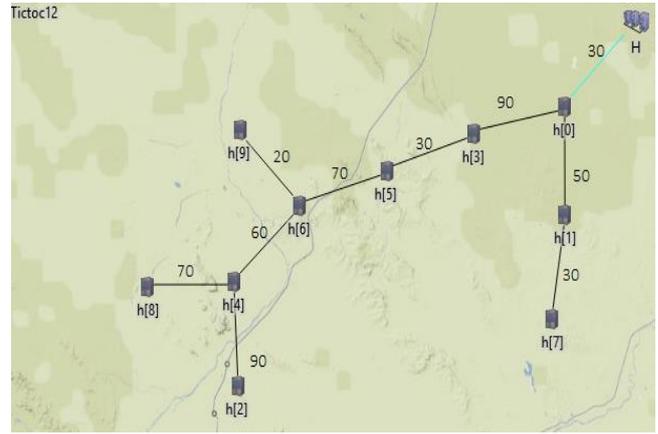

Figure III.1: Sample network of 10 nodes.

1. The managed network will contain a manager node and nodes related to MIB. We assume that the manager is the home node (H), and the computational time at H equals 0.
2. The link agent lists the available node and sorts it in decreasing order based on its routing time, so we can know which nodes need more time to finish the tasks than others. The routing time is calculated as follows (Baek et al., 2001):

tourT(hi) = comphi + 2* Ls(H,hi)

Where tourT(hi) is the time needed to travel from H to the node and back.

The computational time will be the same for all nodes, since we assumed that all nodes send MIB data with a size of 1KB. Calculating routing time for all paths in the available networks takes the path weight into consideration, since the computational time is the same in each node:

a) H0: H→H0 = 30 / tourTh0 = 60
b) H1: H→H0→H1 = 80 / tourTh1 = 160
c) H2: H→H0→H3→H5→H6→H4→H2 : 370 / tourTh2 = 740
d) H3: H→H0→H3 : 120 / tourTh3 = 240
e) H4: H→H0→H3→H5→H6→H4 :280 / tourTh4 = 560
f) H5: H→H0→H3→H5 = 150 / tourTh5 = 300
g) H6: H→H0→H3→H5→H6 = 220 / tourTh6 = 440
h) H7: H→H0→H1→H7 :110 / tourTh7 = 220
i) H8: H→H0→H3→H5→H6→H4→H8 : 350 /tourTh8 =700

j) H9: H→H0→H3→H5 →H6→H9: 240 / tourTh9 = 480

The routes will be ordered in descending order in a list as follows:

H2, H8, H4, H9, H6, H5, H3, H7, H1, H0

3. Let ð be routing time of the first node in the list; in our case, ð = 740.
4. The network will be partitioned based on the routing time so each part should not exceed the value of ð= 740.

In our case the network will be divided into three partitions, and a link agent will generate an agent for each partition, as shown below.

1- Partition 1

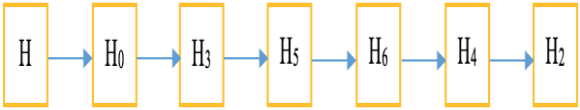

2- Partition 2

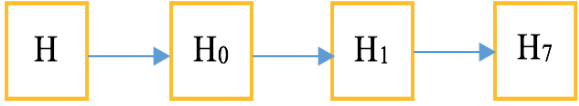

3- Partition 3

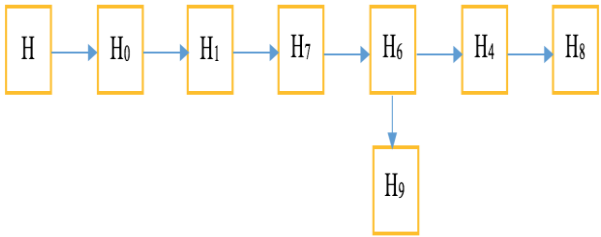

5. Three data agents (A0, A1 and A2) will be sent from H to the three partitions. The tour of each agent is represented in the following tables:

Table 3.1 and Table 3.2 shows the tour of each agent and time spent to retrieve data. Since the size of data at each node is the same - assuming the size of the MIB data that needed to be collected is the same for all nodes and it's equal to 1KB – the first MA is A0, which will be sent to Partition 1. Table 3.1 shows the time taken by A0 to reach the destination node from the home node.

TABLE III.1: A0 time from H to H2.

| Nodes | H | H0 | H3 | H5 | H6 | H4 | H2 |
|---|---|---|---|---|---|---|---|
| Time at node | 0 | 30 | 120 | 150 | 220 | 280 | 370 |

The agent will reach its destination in .0037ms and will need the same time to get back to the home node, thus the A0 tour time is .0074ms.

The second MA, A1, will be sent to Partition 2. Table 3.2 shows the time taken by A1 to reach the destination node from home node.

TABLE III.2: A1 time from H to H7.

| Nodes | H | H0 | H1 | H7 |
|---|---|---|---|---|
| Time at node | 0 | 30 | 50 | 110 |

The agent will reach its destination in .0011ms and will need the same time to get back to the home node, thus the A1 tour time is .0022ms.

The third MA, A2 is sent to Partition 3. Table 3.3 demonstrates the time taken by A2 to reach its destination from the home node.

TABLE III.3: A2 time from H to H8.

| Nodes | H | $H_0$ | $H_3$ | $H_5$ | $H_6$ | $H_4$ | $H_8$ |
|---|---|---|---|---|---|---|---|
| Time at node | 0 | 0.0003 | 0.0012 | 0.0015 | 0.0022 | 0.0028 | 0.0035 |

Since H9 is connected to H6, after running all possible paths to choose the shortest, the agent decided to visit H9 when returning from H8 to the home server. Table 3.4 shows the return rout of A2.

Table III.4: A2 time from H8 to H.

| Nodes | $H_4$ | $H_6$ | $H_9$ | $H_6$ | $H_5$ | $H_3$ | $H_0$ | H |
|---|---|---|---|---|---|---|---|---|
| Time at node | 0.0042 | 0.0048 | 0.005 | 0.0052 | 0.0059 | 0.0062 | 0.0071 | 0.0074 |

As shown in Table 3.3 and Table 3.4, the total routing time for A2 is .0074ms.

From the above, we see that all MAs will have a route of equivalent or less speed than the agent assigned to ð.

The total routing cost is completed at 0.0074s.

## IV. EXPERIMENTAL RESULTS AND MAIN FINDINGS

This section presents a brief about the existing approaches within SNMP to collect MIB, along with the results of the experiments that was conducted on networks with 5, 10,15,20,25, and 30 node regarding number of agents and total routing time at each run. The simulation for the networks build using OMNET++. A comparison is conducted later in this section between our approach and previous approaches presented in (Al-Kasassbeh & Adda, 2008; Al-Kasassbeh & Adda, 2009;Al-Kasassbeh 2011; Wittner, Helvik, & Hoepler, 2000) used within SNMP to retrieve MIB data.

In the traditional SNMP paradigm, each node has its own MIB data, which is obtained by the managed node through management protocols, as shown in Figure 4.1.

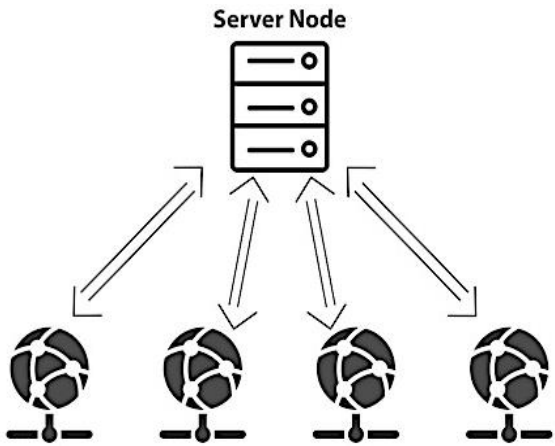

**Figure IV.1:** Traditional paradigm of network managed by SNMP.

This approach is neither reliable nor functional when it comes to a network with large distributed nodes. To overcome this issue, MAs were used in distributed network management, to enhance reliability and flexibility, and reduce network latency. This enhances performance compared to traditional SNMP-MIB approach, by using MAs that are able to move from one node to another for collecting data for network status investigation purposes, as shown in Figure 4.2.

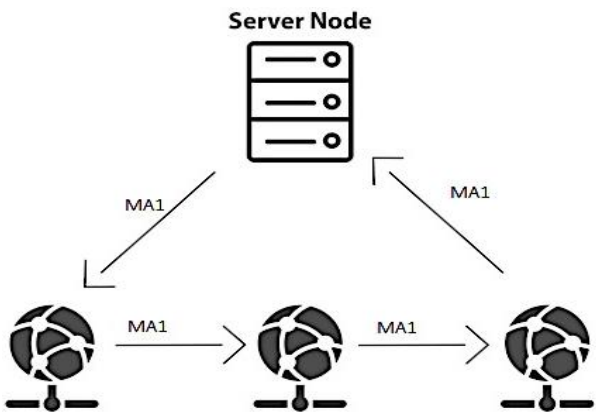

**Figure IV.2:** SNMP with MA.

Al-Kasassbeh and Adda (2008) proposed two approaches that achieve data collection in distributed network using MA:

1- Accumulative Model: MA travels from one node to another, collecting data until the last node in the network, and holding polled data from node to node. This enlarges the size of agents, which affects delays and traffic (Figure 4.2).

2- Interactive Model: in this model the MA reaches the first node, then it clones, and the clone travels to the next node to collect data, and so on. For networks with a large number of nodes, the number of MAs will increase dramatically, as shown in Figure 4.3.

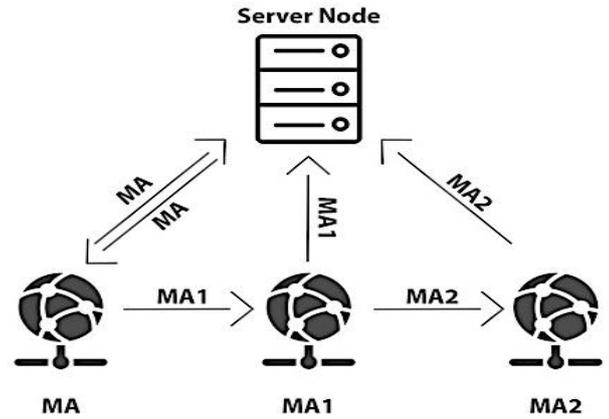

Figure IV.3: Interactive Model for SNMP with MAs.

This paper proposes a dynamic adaptive approach using MAs to collect MIB data from network managed by SNMP as follows:

1- Link MA will run as a manager and listen to the nodes' announcements, to draw the network graph for the live nodes.

2- Nodes in the network advertise nodes to which they are connected, and the weight of each link comes out of the node.

3- Then, Link MA calculates the route weight each node from the home node as follows:

$$\sum_{i=0}^{N} CompHi + 2 * Ls(H, Hi)$$

Where CompHi is the computational time the agent needs to complete a task at Hi, and Ls(H,Hi) is the shortest latency between two nodes (H, Hi)

4- Link agent orders the nodes based on the routs' value in descending order, then the network is partitioned into sub-networks such that the total rout for each one should be less than or equal to the biggest route.

5- Link MA will be cloned based on the number of sub-networks. Each clone is called data agent and will be sent to each sub-network assigned to it.

Table 4.3 summarizes the experimental results regarding number of agents, number of sub-networks, and total routing time in seconds needed to retrieve data at each run.

TABLE IV.1: Results of experiments.

| Number of nodes | Number of sub-networks | MAs | Routing Time (s) |
|---|---|---|---|
| 5 | 2 | 2 | ,0026 |
| 10 | 4 | 4 | ,0066 |
| 15 | 4 | 4 | ,0066 |
| 20 | 6 | 6 | ,008 |
| 25 | 5 | 5 | ,0092 |
| 30 | 8 | 8 | ,0096 |

As per shown above the number of Mobile agents depends on the number of sub-network which is dependent on the routing time of the furthest node. So it's not necessarily that number of MAs increase when number of nodes increase. That means for network with different number of nodes it's possible to send the same number of mobile agents to retrieve MIB data as shown in figure 4.8.

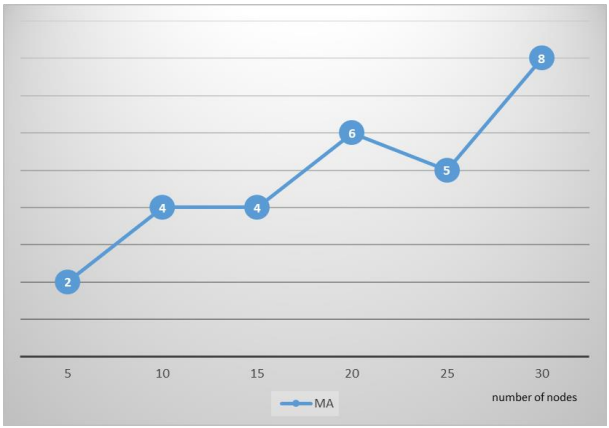

Figure 4.8: MA number regarding to the number of nodes.

In the previous experiments it can be noted that although the number of nodes was increasing, the number of MAs did not increase in all cases; indeed, in some cases the number of nodes was larger, while the number of agents was smaller.

The total routing time was one of the most important aspects we focused on in our algorithm. As mentioned previously one of our goals is achieving the minimum routing cost to pull MIB data that will enhance the process of fault management and detect any threats. Figure 4.9 demonstrates that it is not axiomatic that whenever the node number increases, the time increases as well. Regarding our experiments, the routing cost can be the same even if the number of live node differs at each run.

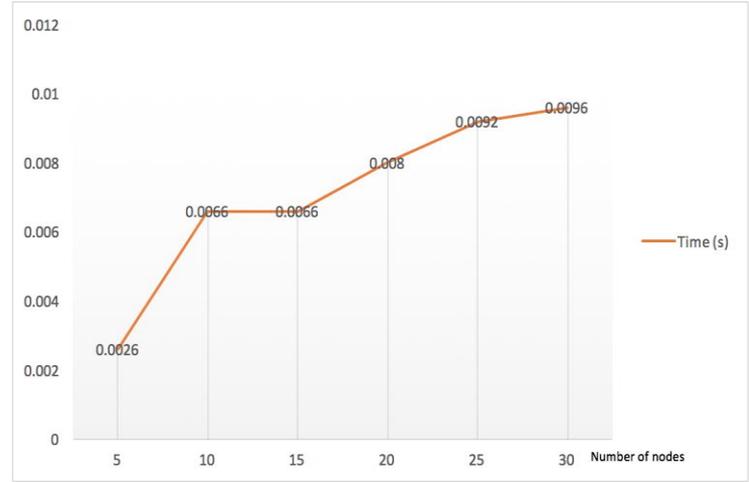

Figure IV.4: Total routing time regarding number of nodes.

Consequently, it can be said that what does affect the number of agents and the time needed to complete that task of gathering MIB data from network is the distance between nodes. Our approach calculations rely on the farthest node, and divide the network into sub-networks, whereby its routing costs do not exceed the routing cost of the furthest node. Unlike Accumulative Model which is using single mobile agent planning and interactive model that is using Multi Mobile Agent Planning but the agent creation relies on the number of the nodes in the network.

Table IV.2 shows the Time differences between Accumulative and Interactive Models and MMAP: Dynamic Time-Effective Approach.

Table IV.3: Time differences between Accumulative and Interactive Models and MMAP: Dynamic Time-Effective Approach.

| Nodes | Time (s) | | |
|---|---|---|---|
|  | Accumulative | Interactive | MMAP-Dynamic |
| 5 | 0.0052 | 0.0052 | 0.0052 |
| 10 | 0.0125 | 0.0078 | 0.0078 |
| 15 | 0.013 | 0.0064 | 0.0064 |
| 20 | 0.0113 | 0.006 | 0.006 |
| 25 | 0.0207 | 0.0078 | 0.0078 |
| 30 | 0.021 | 0.0098 | 0.0098 |

The results shown in Table 4.7 demonstrate that the time in our algorithm and in the Interactive Model is the same, which is because they both depend on the total routing time to reach the farthest node and return to the home node. We can notice that the Accumulative Model's time keeps increasing while the number of nodes increases, because it depends on a single MA, which must visit all nodes in the network. As shown in Figure 4.15, the time for MAs in our algorithm to complete their task is shorter than in the Accumulative Model, but the same as in the Interactive Model (the line for the latter is subsumed in the graph in the MMAP-Dynamic line).

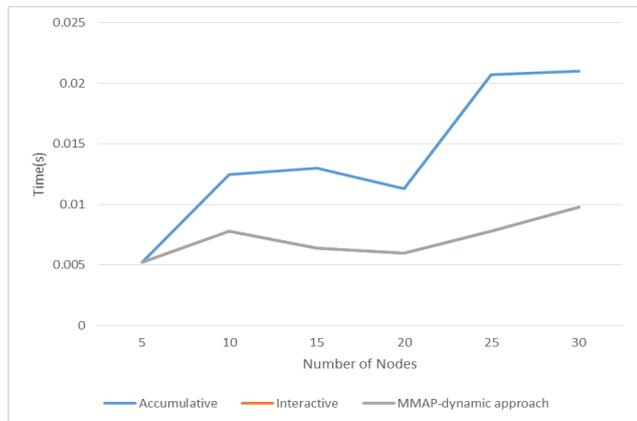

Figure IV.5: **Time difference between three models.**

The identical time of the Interactive and MMAP-Dynamic Models in Figure 4.15 is due to the MAs in both algorithms completing the MIB data collecting task at the same time, which is the same as the routing time to the farthest node in the network. However, we can notice differences between these models in the number of MAs that must be sent through the network to collect MIB data, as shown in Table 4.8.

Table IV.4: Differences between three models regarding number of agents.

| Nodes | Accumulative | Interactive | MMAP-Dynamic |
|-------|--------------|-------------|--------------|
| 5     | 1            | 5           | 1            |
| 10    | 1            | 10          | 2            |
| 15    | 1            | 15          | 4            |
| 20    | 1            | 20          | 6            |
| 25    | 1            | 25          | 5            |
| 30    | 1            | 30          | 6            |

Table 4.8 compares between the Accumulative, Interactive, and the proposed MMAP-Dynamic algorithm. The Accumulative Model uses a single agent to collect all MIB data from the network, regardless of the number of nodes. This is unreliable, since the MA size will increase and performance will decrease.

As shown in Figure 4.16, the number of agents needed to complete the MIB data collecting task in our algorithm is less than the number of agents in interactive algorithm for the same number of nodes, because the number of agents in our algorithm relies on distance and density. This gives our algorithm the capacity to be used in SNMP, since it maintains the number of MAs sent through the network, in contrast with the Interactive Model.

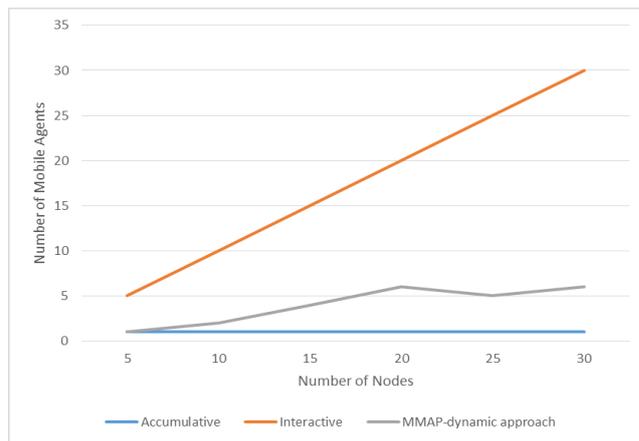

**Figure IV.6:** Differences between three models regarding number of MAs.

Interactive and Accumulative Model were proposed to enhance efficiency and reduce the time needed for Mobile agents' task to collect MIB. Accumulative Model uses single MA to visit all nodes in the network to collect the MIB data then return to the home node. This approach is not effective when it comes to distributed network with a large number of nodes, due to the increase in agent size, leading to an increase in the delay and network traffic, and therefore an increase in the time the agent needs to complete the MIB collecting task.

The Interactive Model aims to overcome the deficiencies of the Accumulative Model by using multiple agents (instead of one) to collect MIB data, such that when first agent reaches the very first node in the network it clones itself, and sends the new MA to the next nodes, to collect their data, and so on until the very last node in the network. When the MA completes its task at the destination node it returns to the home node directly, so the total routing time in this model is equal to the time the agent needs to reach the farthest node and come back.

Thus, a network having $N$ number of nodes needs $N$ MAs to collect its MIB data, therefore for networks with very large nodes, a very large number of agents is disseminated to collect MIB data, which increases overheads over network. Our approach overcomes the problem of the number of travelled MAs in the Interactive node and keeps traffic at a minimal level, increasing the additivity of MAs to the network changes and ensuring that the time agents need to complete their task is minimal.

Our approach relies on two types of agents; link agent and data agent. Link agent is the agent that runs our algorithm and decides how many agents need to be sent to the network to collect the MIB data. Our algorithm divides the network so that the total time of each sub-network does not exceed the time needed to reach the farthest node. A single MA is then sent to each part to collect MIB data and return after the MA visits all nodes in the assigned sub-network. This means that for a network with $N$ nodes the number of agents needed to be sent for MIB data collection purpose is defined based on number of sub-networks, which in turn depends on the value of the total routing time for the furthest node in the network, making our algorithm more reliable, efficient, and adaptive than the traditional paradigm, the Interactive and Accumulative Models.

V. CONCLUSION

This section summarizes the main accomplishments of this research, summarizes the observed results, and outlines future research directions arising from these findings.

*A. Outcomes*

Using network management systems protects the network from attacks like DoS, because it works as an IDS. SNMP is one of the most well-known and most widely implemented systems in network devices. MIB is a tree structured database that contains data usable for detecting malicious activities and anomalies.

Using decentralized MA planning in the distribution system is not effective due to latency and delay. It is also not adaptive to network changes related to the availability of the nodes and the densities.

A multi MA planning model was used in the distribution system to find the optimal number of MAs to complete an information retrieval task with the minimum routing time, with minimal impacts on network traffic.

This paper uses MMAP to retrieve MIB data with minimum routing time and cost, without significantly affecting the network bandwidth. This enhances IDS responsivity to attacks.

Our algorithm used two type of agents: link and data agent. The link agent is responsible for discovering the network and storing the connected nodes in the home server. The data agent is responsible for the data retrieval task. The nodes that are assigned to agents were determined after partitioning the available nodes based on routing time.

Agents were sent to each part of the network. Our algorithm was compared to two previous models serving the same goal (the Interactive and Accumulative Models). The results prove that our model is better for application in distributed network with a large number of nodes in terms of the number of agents and time, since it is not affected by the increasing node number, unlike the alternative models.

*B. Recommendations and Future Work*

In future work we want to enhance to this algorithm and enhance the cloning method so it can be conducted on the top of a tree in the network. This means sending an agent to nodes with similar paths, and letting the cloning happen at the points to which the network branches out, to reduce number of agents thus reduce the traffic when it comes to networks with a large number of nodes.